\documentstyle[epsfig,pra,aps,preprint]{revtex}
\tightenlines
\begin{document}
\title{\bf
Adiabatic stabilization of a circular state: theory compared to experiment \\}

\author{Bernard Piraux$^1$ and R.\ M.\ Potvliege$^2$}

\address{\mbox{}$^1$Institut de Physique, Universit\'e Catholique de Louvain,
2 Chemin du Cyclotron, B-1348 Louvain-la-Neuve, Belgium \\
\mbox{}$^2$Physics Department, University of Durham, Science Laboratories,
Durham DH1 3LE, England}

\maketitle

\begin{abstract}
{The probability that an atom of hydrogen, initially in the $5g(m=4)$ state,
survives a 90-fs pulse of 620 nm wavelength is calculated both by direct
integration of the time-dependent Schr\"odinger equation and by a
Floquet calculation. The two methods give virtually identical results.
The survival probability calculated for a one-electron model
of neon, for the same initial state, pulse duration and wavelength,
is in fair quantitative agreement with the experimental data of
van Druten {\it et al.} [Phys.\ Rev.\ A {\bf 55}, 622 (1997)].}
\end{abstract}

\begin{center}
To be published in Physical Review A, June 1998
\end{center}

\narrowtext

As is well known, Fermi's golden rule
predicts that an atom irradiated by a weak laser field
decays by photoionization at a rate proportional to the intensity of the field
if the photons energy exceeds the binding energy. This simple
perturbative result is true, however, only at low intensity. It is now
widely accepted that the rate of ionization in an intense high-frequency laser
field increases {\em less} rapidly with intensity than
a linear power law, and may even {\em decrease} above a certain critical
intensity (before increasing again at ultra-high intensity when the coupling
with the magnetic field of the incident wave becomes important).
This remarkable reduction in the photoionization rate is often referred to
as ``adiabatic stabilization",
even though the atom never becomes completely stable against photoionization,
because the quasi-stationary wave packet describing the atom at high intensity
can be traced adiabatically to its field-free, initial state; in particular,
its formation does not depend critically on fast variations of the
intensity \cite{HGM}.

Adiabatic stabilization
was first studied in depth in the late 1980s, by M.\ Gavrila,
M.\ Pont and co-workers \cite{Gavrila}, and subsequently by many others.
Numerical calculations focused initially on the stabilization of the
ground state or low-excited states of hydrogen or 
model systems \cite{1990}. These were extended in 1991 and later
to the analysis of atoms initially in an
excited state with a high
magnetic quantum number with respect to the polarization direction
\cite{circ}. While results interpreted by several authors as casting doubts
on the existence of the effect were also reported \cite{Geltman},
the numerical calculations in general showed that the ionization rate indeed
decreases at sufficiently high intensity. In particular, the calculations
for atoms initially prepared in an excited circular state
($|m| = l \approx n-1 \gg 1$) demonstrated
stabilization
at intensities, frequencies and pulse durations that could be achieved
in the laboratory.
This prompted two experiments --- the 
only experimental test of adiabatic stabilization so far --- in which the
photoionization of the $2p^55g(m=4)$ state of neon was studied
at a wavelength of 620 nm and intensities up to $2.3 \times 10^{14}$ W/cm$^2$
\cite{deBoer,vanDruten}. Onset of
stabilization was observed at high intensity, and,
to the extent that a comparison could be made, the experimental results were
found to be consistent with the results of
calculations done at the same wavelength for the $5g(m=4)$ state of hydrogen.

The present work aims at a closer comparison of theory with the
experimental data of Ref.\ \cite{vanDruten}.
The quantities measured in the experiment (the yield in photoelectrons and
the residual populations at the end of the pulse) can be calculated, for
a simplified model of the atom,
either by direct integration of the time-dependent Schr\"odinger equation
or by treating the system as being quasi-stationary and using Floquet methods.
We first show that, in the present case,
these two approaches give virtually identical results.

Both the Floquet and the time-dependent calculations were performed in the
dipole approximation, with the field described by the vector potential
${\bf A}(t) = A_0(t) \sin \omega t \, {\bf \hat z}$.
For $A_0$ constant and with
Siegert boundary-conditions imposed on the wave function,
the quasienergy of the (Floquet) dressed state, $E$,
is a complex eigenvalue of a system of time-independent differential equations.
We solved this system numerically in the velocity gauge, using 
a basis of spherical harmonics and complex radial Sturmian functions
\cite{method}.
The quasienergy was also obtained by Pad\'e-summing its Rayleigh-Schr\"odinger
perturbative expansion in powers of the intensity. In the high frequency
regime, the latter technique gives the same results as a full non-perturbative
calculation for a small fraction of the cost in CPU time.
The rate of photoionization,
$\Gamma = -2 \, \mbox{Im }(E/\hbar)$, does not depend on the 
normalization of the Floquet wave function and 
is a function of the vector potential amplitude, $A_0$, and of the other
parameters of the incident field.
The variation of the ionization rate with intensity
is shown in Fig.\ 1, for the dressed $5g(m=3)$ and $5g(m=4)$ states
of hydrogen at 620 nm wavelength:
$\Gamma$ increases in proportion of the intensity in weak fields,
in agreement with the golden rule, but eventually reaches a maximum and then
decreases monotonically as stabilization sets in.
The maximum occurs at a higher intensity for
$m=3$ than for $m=4$, as is discussed in Ref.\ [4d].
Note that in Fig.\ 1,
$\Gamma$ remains small compared to the
inverse of a laser period, $\omega / 2 \pi = 0.012$ a.u., and can therefore
be interpreted in the usual sense, as a cycle-averaged ionization rate.
The process being non-resonant, one may expect that the atom follows
the dressed states adiabatically if $A_0$ varies slowly instead of being
constant. Our Floquet calculations thus relied on representing the
atom by a single
dressed state, i.e.\ the one that develops from the unperturbed initial state
as $A_0$ increases. Accordingly, the probabilities
that the atom survives a (not too short) laser pulse and is left in its
original state, $P_{\rm surv}$, or that it is photoionized,
$P_{\rm ion}$, were taken to be
\begin{equation}
P_{\rm surv} =
 \exp \left( -\int_{-\infty}^{+\infty}\Gamma[A_0(t)] \, {\rm d}t \right)
\end{equation}
and $P_{\rm ion} = 1 - P_{\rm surv}$
(thus neglecting excitation of other dressed states).

The method used for numerically integrating the time-dependent
Schr\"odinger equation, for atomic hydrogen,
has been described in Ref.\ \cite{methodTD}. Briefly,
the wave function, represented on a basis of spherical harmonics
and (real or complex) radial Sturmian functions,
is propagated in time using a very stable
predictor-corrector algorithm based on implicit Runge-Kutta methods.
The photoionization yield is obtained as the complement to unity of the
population remaining bound to the nucleus at the end of the pulse, which
is calculated by projection on all atomic bound states with $n \leq 25$
and $l \leq 18$. When using real Sturmians,
the basis must span a region of space large enough to contain the wave packet
until the end of the pulse. Much smaller bases are
sufficient when employing complex Sturmians, as they smoothly absorb 
the outgoing flux before it leaves the region where the atom is
represented accurately.
The results quoted below have been obtained by using complex 
bases, of typically 15 spherical harmonics (with $\ell$ varying from 4 to 18)
and, for each partial wave, 120 complex radial Sturmians.
We have verified that at high field intensity,
$2 \times 10^{14}$ W/cm$^2$, identical results (to three significant 
figures) are obtained with
a large basis of 600 real Sturmians per angular momentum and 50 angular 
momenta.

In order to compare the Floquet and the time-dependent methods, we calculated
the probability that an hydrogen atom, initially in the $5g(m=4)$ state with
respect to the polarization direction, is ionized by a 120-cycle pulse.
The pulse duration (90 fs, FWHM in intensity) was the same as in the experiment.
For convenience, its envelope was taken to be $A_0(t) = A_0\sin^2 (t / \tau)$
\cite{Schrader}.
The results are shown in Table I. Only insignificant 
($< 0.0001$ \%) residual populations in bound states other than
the $5g$ state were found in the time-dependent calculation.
Adiabatic stabilization manifests by 
a {\em decrease} in the ionization probability above about
$1 \times 10^{14}$ W/cm$^2$ peak intensity. The decrease is less pronounced in
the ionization probability than in the ionization rate, but is
still remarkable.
Also noteworthy is the 4-digit agreement between the Floquet results and the
time-dependent results. The agreement is better than had been previously
reported for much shorter pulses \cite{timedep}. It confirms that the one-state
Floquet approximation is reliable for describing ionization in high
frequency pulses, and makes it possible to assign, unambiguously,
the reduction in the ionization probability
observed in the time-dependent results to adiabatic stabilization \cite{BandR2}.

The probability that the atom remains in the original $5g(m=4)$ state is
also the same in the two methods, since
there is no net population transfer to other bound states.
This is illustrated by Fig.\ 2. The figure
also gives the survival probability for a pulse of same width as
above but with a sech$^2$ intensity profile: the probability is a little smaller
for the sech$^2$-pulse than for the finite-duration pulse because
for the same fluence the former irradiates the atom
at (relatively) weak intensities for a longer time compared to the latter.
A simple calculation shows that for a fixed pulse length and increasing fluence,
the survival probability tends asymptotically to a constant smaller than 1 for
a sech$^2$-pulse,
assuming of course adiabaticity and a continuing decrease of $\Gamma$ at
high intensity. However,
it decreases to zero for pulses with, for example, a Lorentzian profile or
increases to 1 for finite-duration pulses.
Thus adiabatic stabilization (the decrease in the ionization rate)
does not necessarily translate into a decrease in the ionization probability.

We now compare our theoretical results with the experimental data.
We model the neon atom by a single electron moving in a central potential,
$V(r)$,
that accounts for the partial screening of the nuclear attraction by the
inner electronic shells. In atomic units \cite{Feneuille},
\begin{equation}
V(r)= -{1 \over r} [2 f_0(\alpha_1 r) + 2 f_0(\alpha_2 r) + 5 f_1(\alpha_3 r)
 + 1]
\end{equation}
where the parameters $\alpha_1 = 13.06$, $\alpha_2=7.2$ and $\alpha_3=3.68$
were obtained by fitting the energy levels of $V(r)$
to experimental values, and
\begin{equation}
f_\ell(\alpha r) = e^{-\alpha r} \sum_{j=0}^{2\ell+1}
\left(1 - {j \over 2\ell+2}\right) {(\alpha r)^j \over j!}.
\end{equation}
Not surprisingly,
the short-range part of $V(r)$ is of very little importance for states
with $m=3$ or 4. The rates of ionization from the dressed $5g(m=3)$ and
$5g(m=4)$ states of the model atom are within 0.02 \% and 0.0005 \%,
respectively, from those based on a pure Coulombic potential
(shown in Fig.\ 1).

The fraction of the atomic population left in the initial state at the end
of a 90-fs pulse is shown in Fig.\ 3. The surviving
fraction was measured directly in the experiment, and was also estimated
indirectly as the complement to unity of the measured
ionized fraction. The two sets of data are consistent
and do not follow the rapid decrease at high fluence that
a naive application of the
golden rule would predict (see the dotted line in the figure, or Fig.\ 8 of
Ref.\ \cite{vanDruten}).
The solid line in Fig.\ 3 represents the result of the Floquet
calculation for the sech$^2$-pulse of above.
We assumed a uniform distribution of intensity
since in the experiment the target atoms were prepared in the initial
excited state only in the central part of the 620 nm beam.
We also assumed that the system was initially in a pure $5g(m=4)$ state
while in the experiment the atoms were initially in a superposition of
several correlated states with a small (at most 13 \%) admixture
of states having a $5g(m=3)$ character. The importance of this admixture 
was tested by repeating the calculation for an
incoherent superposition of the $5g(m=4)$ and $5g(m=3)$ states in a
87~\%~/~13~\% ratio. The resulting survival probability, represented by
the dashed line, is smaller than for the pure $5g(m=4)$ state, since
ionization proceeds faster from the $m=3$ state, but the difference is minor.
As seen from the figure, the calculated probabilities and the data are in 
fair quantitative agreement. This, of course, tails well with the
conclusion of van Druten {\it el al.} \cite{vanDruten},
that the large survival probability
at high fluence observed in the experiment
can be attributed to adiabatic stabilization.

The authors thank J.\ F.\ McCann for his critical reading of the
manuscript. B.P.\ is supported by the
``Fonds National de la Recherche Scientifique".

\begin{table}
\caption{Probability that an atom of hydrogen, initially in the $5g(m=4)$ state
is ionized by the 120-cycle laser pulse described in the text, as calculated
either by full numerical integration of the time-dependent Schr\"odinger
equation or by Floquet methods. The wavelength is 620 nm and the polarization
is linear. The reduced mass of the system was taken to be 1 a.u.\ in both
calculations.}
\begin{tabular}{lcc}
Peak Intensity  & Time-dependent & Floquet \\
\hline
$0.25 \times 10^{14}$ W/cm$^2$     & 0.10103 & 0.10105 \\
$0.50$                             & 0.14122 & 0.14126 \\
$1.00$                             & 0.16035 & 0.16042 \\
$1.50$                             & 0.15596 & 0.15599 \\
$2.00$                             & 0.14705 & 0.14705 \\
\end{tabular}
\end{table}
\begin{figure}[h]
\centerline{\epsfig{figure=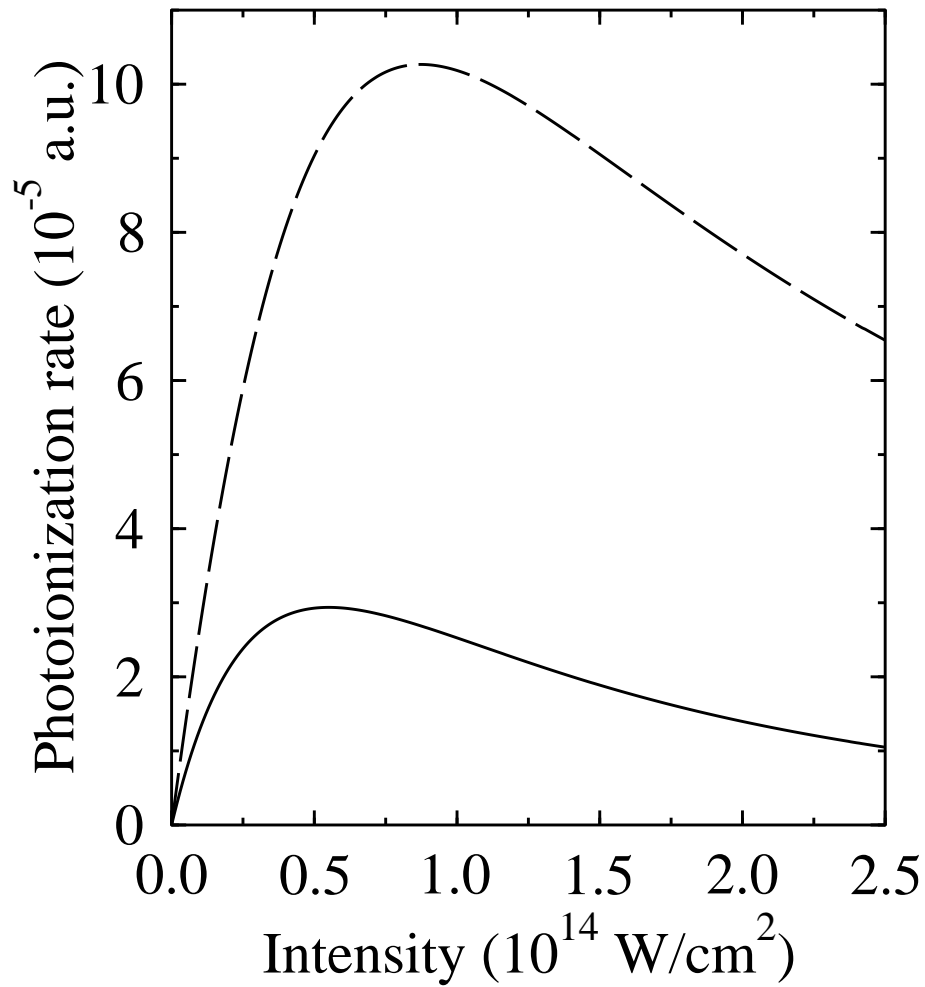,width=0.95\hsize}}
\caption{Variation with intensity of the
rate of photoionization of atomic hydrogen from the $5g(m=3)$ state
(dashed line) or the $5g(m=4)$ state (solid line) at 620 nm wavelength.
The field is linearly polarized along the axis
of quantization of the angular momentum.
The rate is expressed in atomic units.
}
\end{figure}
\begin{figure}[h]
\centerline{\epsfig{figure=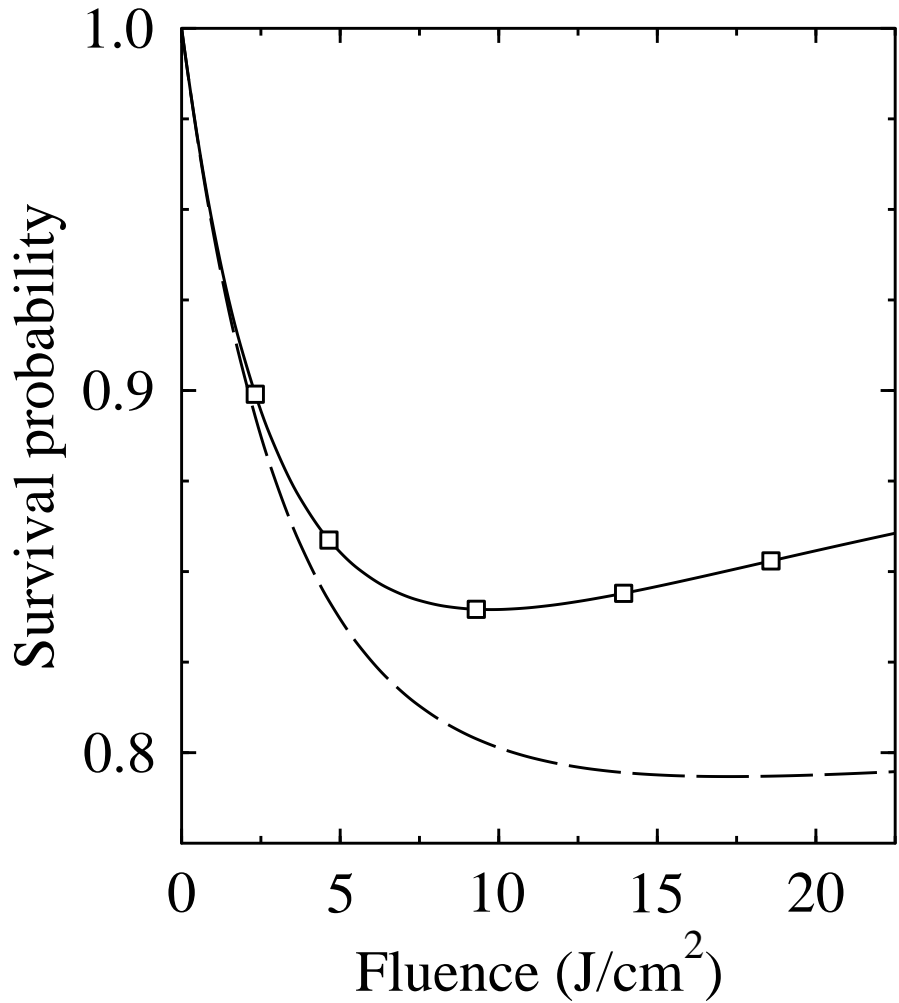,width=0.95\hsize}}
\caption{
Probability that an atom of hydrogen initially in the 
$5g(m=4)$ state of hydrogen is in the same state at the end of
a pulse of 620 nm wavelength and 90 fs duration (FWHM in intensity),
as a function the pulse fluence. 
The field is linearly polarized along the axis
of quantization of the angular momentum.
Solid line: results of the Floquet calculation for
the finite-duration pulse defined in the text.
Squares: results of the time-dependent calculation for
the finite-duration pulse.
Dashed line: Floquet results for a pulse with a sech$^2$-profile
in intensity.
}
\end{figure}
\begin{figure}[h]
\centerline{\epsfig{figure=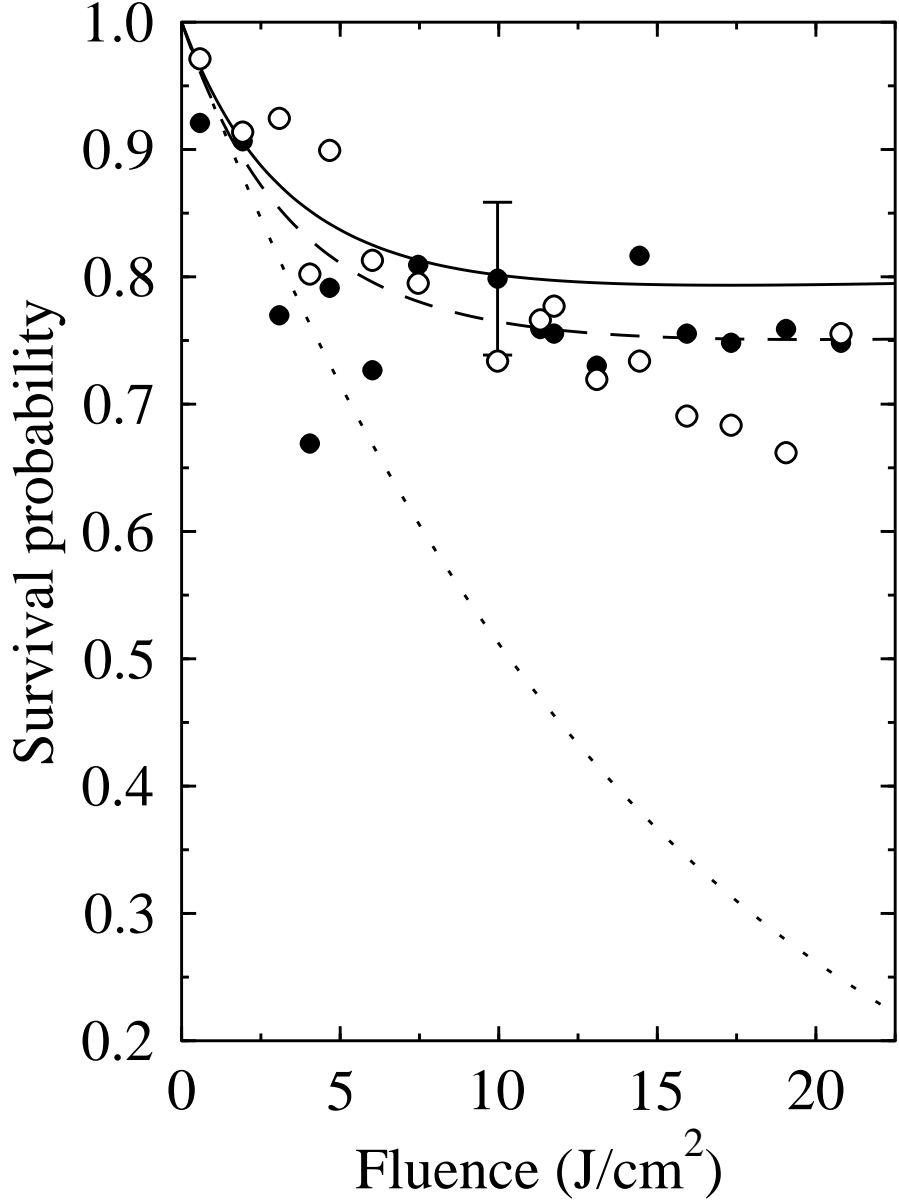,width=0.95\hsize}}
\caption{
Surviving fraction of the $2p^55g$ population.
Open circles: surviving fraction measured by van Druten {\it et al.}
[7].
Solid circles: complement to unity of the ionized fraction
[7].
Solid curve: probability calculated non-perturbatively for
a pure $5g(m=4)$ initial state.
Dashed curve: same as the solid curve, but for an incoherent superposition of 
$5g(m=4)$ state (87 \%) and $5g(m=3)$ state (13 \%).
Dotted curve: prediction of first-order perturbation theory for a
pure $5g(m=4)$ initial state.
}
\end{figure}
\end{document}